\documentclass[prl,twocolumn,showpacs,preprintnumbers,amsmath,amssymb]{revtex4}
\usepackage{graphicx}% Include figure files
\usepackage{dcolumn}% Align table columns on decimal point
\usepackage{bm}% bold math
\begin{document}

%\preprint{APS/123-QED}

\title{Brillouin Lasing with a CaF$_2$ Whispering Gallery Mode Resonator}

\author{Ivan S. Grudinin}
\altaffiliation[Also at]{Physics Department, California Institute of Technology.}
\email{grudinin@caltech.edu}
\author{Andrey B. Matsko$^{\dag}$}
\author{Lute Maleki}
\altaffiliation[Present address ]{Oewaves Inc., 1010 East Union Street, Pasadena, CA 91106}
\affiliation{Jet Propulsion Laboratory, California Institute of Technology, \\ 4800 Oak Grove Drive, Pasadena, CA 91109-8099}
\date{June 20, 2006}% It is always \today, today,
\begin{abstract}
Stimulated Brillouin scattering with both pump and Stokes beams in resonance with whispering gallery modes of an ultra high Q $CaF_2$ resonator is demonstrated for the first time. The resonator is pumped with 1064\,nm light and has a brillouin lasing threshold of $3.5\ \mu W$. Potential applications include optical generation of microwaves and sensitive gyros.
\end{abstract}
\maketitle

Stimulated Brillouin scattering (SBS), which produces a reflected Stokes beam in a fiber ring resonator, may be viewed as a lasing process. In this case, scattering of light by an acoustic wave provides gain. While scattering may be a power limiting factor in fiber communications, Brillouin lasers have several attractive features. Many Stokes components may be generated in a cascade, with thresholds below 1\,mW. Simultaneous detection of these components may generate microwaves in 10--100\,GHz range. An interesting feature of the fiber Brillouin lasers is the linewidth narrowing effect, where the Stokes line may be narrower than the pump laser line by a factor of up to $10^4$. Relative intensity noise is also reduced in these lasers. Thus applications of Brillouin fiber ring lasers \cite{geng06ieeeptl} include rotation sensing \cite{zarinetchi91ol}, linewidth narrowing, microwave frequency generation and high rate amplitude modulation.

Ultra--high Q whispering gallery mode (WGM) resonators \cite{matsko06jstqe} enhance the efficiencies of nonlinear processes by combining strong spatial confinement of light with long effective optical path. This results in a significant decrease of threshold for regular lasing, stimulated Raman scattering \cite{grudinin07ol} and other nonlinear processes \cite{delhaye07nat}.

Advantages of WGM resonators and Brillouin lasers imply superb characteristics for a WGM Brilloin laser. In particular, the linewidth narrowing depends on both optical and acoustic quality factors of the resonator \cite{geng06ieeeptl}:
\begin{equation}
\Delta\nu_{Stokes}=\frac{\Delta\nu_{pump}}{(1+\frac{\pi\gamma_B}{\Gamma_C})^2},
\end{equation}
where $\gamma_B$ is the Brillouin gain bandwidth and $\Gamma_C$ is the optical loss rate of a cavity.
Brillouin gain bandwidth in fluorite material is 12.2\,MHz \cite{sonehara07josab}, and resonators with quality factor on the order of $10^{11}$ were recently demonstrated in our lab. Thus linewidth narrowing factors on the order of $2\times10^8$ may be possible. If the pump laser linewidth is 5\,kHz, the expected Brillouin laser linewidth is just $30\ \mu Hz$. Mixing two Stokes from the same cavity could produce microwave signal with narrow linewidth $\Delta\nu_{microwave}=\sqrt{\Delta\nu_{B1}^2+\Delta\nu_{B2}^2}=42\ \mu Hz$.
Theoretical analysis shows \cite{savchenkov07josab}, that a laser locked to a crystalline cavity may have good stability. Hence a crystalline WGMR based Brillouin laser could be used for compact optical frequency standards, sensitive gyro and microwave generators. It should be noted, however, that at room temperature the linewidth of the Stokes component will be limited by noises caused by temperature fluctuations in the resonator material. If SBS were to be observed in fused silica WGM resonators, the linewidth reduction ratio would be limited to $3\times10^6$ by the maximum achieved optical Q factor. Silica resonators also require isolation from the air and do not support spectrum tailoring, which is available for crystalline resonators.

Even though bulk Brillouin gain generally exceeds Raman gain, SBS is generally believed not possible in the ultra--high Q solid state microresonators. There would be no WGM present at a Brillouin offset to support the Stokes beam. The SBS Stokes offset is generally around tens of GHz and the gain bandwidth is only a few tens of MHz. While the free spectral range (FSR) of WGM microresonators is typically in the hundreds of gigahertz. In addition, the acoustic frequency has to be resonant with one of the mechanical modes of the microresonator, otherwise the acoustic wave interferes destructively with itself. Finally, the pump mode, the Stokes mode and the acoustic mode must have a non-vanishing overlap integral. Modes with close frequencies in microresonators generally do not overlap in space. In crystals, speed of sound depends on direction and polarization of the wave, making it harder to fulfil the resonant conditions in crystalline microresonators. This situation is alleviated in either fiber ring resonators, where FSR is comparable to the Stokes offset, or in low Q liquid microdroplets where optical resonance covers both pump and Stokes frequencies \cite{zhang89josab}. Microdroplets, however are not fiber compatible, while fiber ring resonators are large.

In this paper we show that SBS can occur in ultra--high Q millimeter--scale WGM resonators. In such resonators FSR becomes comparable to the Brillouin Stokes shift.  We present the first ever observation of an efficient SBS in a CaF$_2$ WGM resonator and show that SBS can be either enhanced or suppressed by manipulating the modal structure of the optical resonator.

\begin{figure}[htb]
\centerline{\includegraphics[width=8cm]{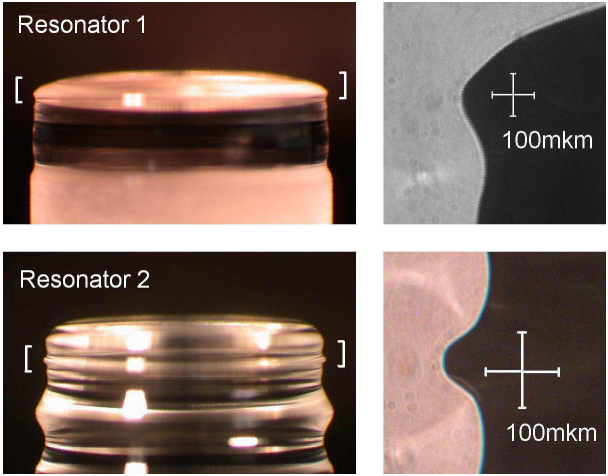}}
\caption{\label{fig:diskphoto} Fluorite resonators. WGM localization areas are highlighted with the white brackets on the left. Shadow photographs of the highlighted regions are on the right. Diameters are 5.52\,mm for cavity 1 and 5.21\,mm for cavity 2.}
\end{figure}

Two ultra high Q single-crystal CaF$_2$ WGM resonators, shown in Fig.\,\ref{fig:diskphoto}, were fabricated. Cavity 1 was fabricated with a UV-grade fluorite from Edmund Optics and had a larger radius of curvature as compared to the cavity 2, which was made with an excimer--grade fluorite from Corning. A Nd:YAG laser with 5\,kHz linewidth manufactured by Lightwave Electronics was used to excite WGMs at the wavelength of 1064\,nm. The diameters of the resonators were determined from the FSR highlighted by multiple Raman Stokes lines observed at 1102\,nm. The spectra of the resonators contain an abundance of WG modes with the average spacing of 25\,MHz for both polarizations.
The laser frequency was locked to a selected cavity mode using a simplified Pound--Drever--Hall technique. A signal generator was used to provide the excitation for the resonant phase modulator. This signal was also used to synchronize the lock--in amplifier SR844 which operated as a frequency mixer and servo. This technique made it possible to maintain a fixed optical power in the cavity.

\begin{figure}[htb]
\centerline{\includegraphics[width=8cm]{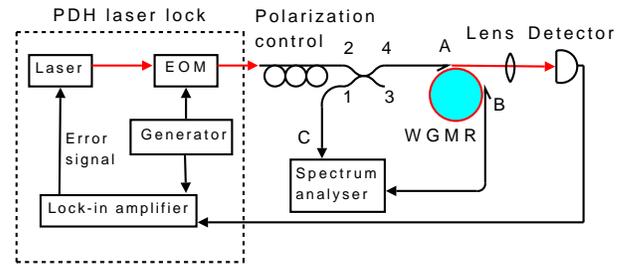}}
\caption{\label{fig:setup} Experimental setup diagram. Laser is locked to a cavity mode. Optical power in the fiber can be monitored in forward and backward directions.}
\end{figure}

\begin{figure}[htb]
\centerline{\includegraphics[width=10cm]{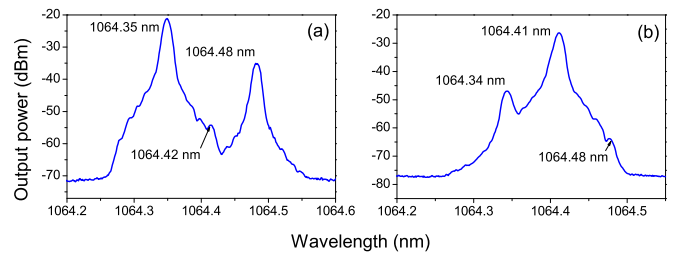}}
\caption{\label{fig:disks} a) Output of the cavity 1, coupler B. Low power at the spectrum analyzer is due to undercoupling. Frequency shift is 17.5\,GHz for the weak Stokes and 35\,GHz for the strong Stokes line. b) Backward signal from output C for second resonator. Stokes offset is 18\,GHz for the first and 35.5\,GHz for the second line. The width of each line is limited by the optical spectrum analyzer resolution of 0.012\,nm.}
\end{figure}

\begin{figure}[htb]
\centerline{\includegraphics[width=8cm]{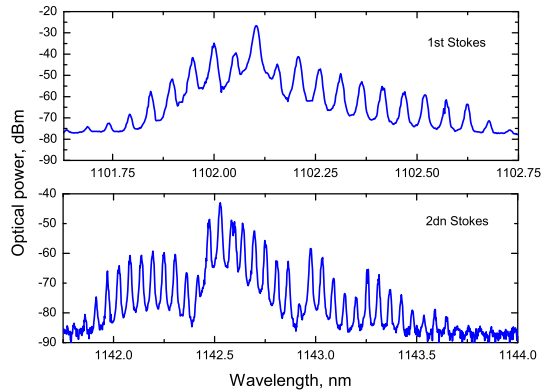}}
\caption{\label{fig:raman} Cascaded Raman lasing observed at pump power above $10\ \mu W$ in the second cavity. Intensity modulation resembles Brillouin spectrum near the pump wavelength.}
\end{figure}

\begin{figure}[htb]
\centerline{\includegraphics[width=8cm]{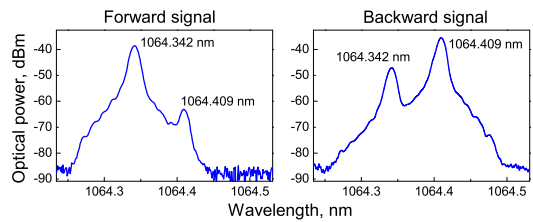}}
\caption{\label{fig:backforward} Typical optical spectra from couplers B and C, representing forward and backward signals. Stokes offset is 17.7 GHz.}
\end{figure}

Alignment of the angle-polished fiber couplers mounted onto the three--axis piezo positioning stages provided input and output coupling efficiencies of up to 80\%. The light that was scattered backwards by the WG mode was branched into the arm of a 90/10 fiber coupler. The setup schematically shown in Fig.\,\ref{fig:setup} was used to measure the power and spectral properties of both forward and backward beams with a Yokogawa AQ6319 optical spectrum analyzer (OSA) and a Thorlabs detector DET10C.
A WGM was excited with approximately $50\,\mu W$ of optical power in a measurement conducted with the first resonator. The spectrum of light from coupler B (Fig.\,\ref{fig:disks}-a) contains a Stokes line offset by 34.9\,GHz and a weak Stokes line offset by 17.5\,GHz. Neither stimulated Raman scattering (SRS) nor four-wave mixing (FWM) oscillations were observed in this measurement. When the pump power was further increased, two groups of Raman lines appeared around 1101.8--1102.3\,nm and 1142.4--1142.9\,nm.
A set of similar measurements was made with the second cavity. The excited WGM had a coupling efficiency of 70\%, loaded Q factor of $4\times10^9$ and an intrinsic Q of around $1.4\times10^{10}$. The spectrum of the backscattered light is shown on Fig. \ref{fig:disks}-b.
To determine threshold, pump power was reduced to $2.9\,\mu W$ below which the first Stokes line was still present but weak. Above this power the intensity of the Stokes line jumps by around 10\,dB and grows quickly with further pump increase. Hence, the threshold of the SBS is around $3\,\mu W$. At pump power well above $10\,\mu W$ Raman lasing was observed in the second cavity as well. The Raman spectrum is shown on Fig.\,\ref{fig:raman}.
For the second cavity we also carried out simultaneous measurements of the forward and backward optical spectrum using couplers B and C (Fig.\,\ref{fig:backforward}). The asymmetry of the spectra may be explained by the residual power leakage between inputs 2 and 1 of the 90/10 coupler, increasing the backward signal at the pump wavelength.
The spectra of our resonators had no observable resonance splitting, which means that only a minor amount of Rayleigh scattering is present for any given WGM. The excitation of a WGM at 1064\,nm created a Stokes line offset by 17.5\,GHz, which propagated in the direction opposite to the pump, as seen on the Fig.\,\ref{fig:disks}-b. This backscattered beam creates a cascaded Stokes component red-shifted by another 17.5\,GHz propagating in the same direction as the pump beam (Fig. \ref{fig:disks}-a). A small feature at the double offset on Fig.\,\ref{fig:disks}-b may be explained as Rayleigh scattering from the second Stokes line visible at Fig. \ref{fig:disks}-a. The weak feature at Fig.\,\ref{fig:disks}-a may also be understood as Rayleigh scattering from the first Stokes line seen at Fig.\,\ref{fig:disks}-b.  These two Stokes lines modify the Raman gain in the resonator, which leads to the SRS spectrum with modulated intensity as seen in the Fig.\,\ref{fig:raman}. Given the OSA resolution of 0.012\,nm and the wavelength repeatability of 0.002\,nm, or 1\,GHz, we conclude that the two offsets of the observed Stokes lines are multiples of 17.7\,GHz. This corresponds to Brillouin scattering by longitudinal phonon branch of fluorite, as we show below.

\begin{figure}[ht]
\centerline{\includegraphics[width=8.5cm]{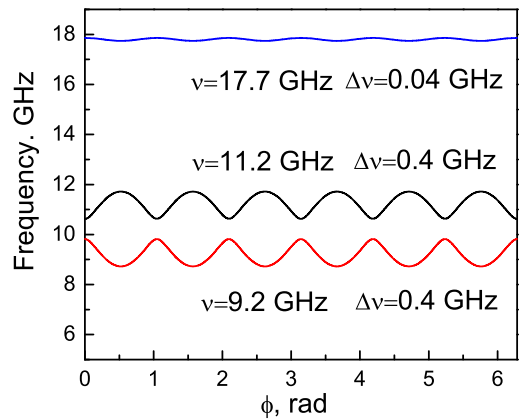}}
\caption{\label{fig1}  Characteristic frequency of the acoustic wave as a function of the azimuth angle in a resonator. }
\end{figure}

Let us compute the Brillouin offset in a CaF$_2$ WGM resonator. We model the WGM as a toroid with axis along the (1,1,1) crystalline orientation. The phonon lifetime $\tau_{ph} \approx 0.08\ \mu s$ is inverse of the spectral width of the SBS gain, which is $12.2$\,MHz in fluorite \cite{sonehara07josab}. The roundtrip time of the sound wave in the resonator is $2\pi a/V_s = 2\ \mu s$. Thus phonons do not form standing waves and SBS in millimeter--sized resonators is similar to scattering in bulk. To find the frequency offset of the SBS we need to know the direction of the light wave vector in the WGM with respect to the crystalline axes. The propagation direction of the electromagnetic wave is given by the following vector:
\begin{equation}
{\bf k} = \left ( \frac{\cos \phi }{\sqrt 6}-\frac{\sin \phi }{\sqrt 2}, \frac{\cos \phi }{\sqrt 6}+\frac{\sin \phi }{\sqrt 2},   -\sqrt{\frac 2 3} \cos \phi  \right ),
\end{equation}
where $\phi$ is the azimuth angle, representing different points along the resonator circumference. The geometry of our system allows back scattering only, so the Stokes beam has a wave vector
${\bf k'} \simeq -{\bf k}$.
The wave vector for the phonons participating in the scattering is ${\bf q}={\bf k'}-{\bf k}$, so the unit wave vector of the sound is ${\vec \kappa} = -{\bf k}$.
CaF$_2$ is a cubic crystal described by three independent elastic constants. From the specifications of our fluorite sample we have: $C_{11}=1.6420$, $C_{12}=0.4398$, $C_{44}=0.3370$ [$\times 10^{11}$ Pa]. Substituting these values into the set of equations presented in \cite{fabelinskiibook} we numerically find the frequency shifts of the Brillouin Stokes components as a function of angle $\phi$ for three different elastic waves (see Fig. \ref{fig1}). Computations were carried out in ``Mathematica'' and ``Maple'' software packages.
Only one solution has a frequency shift that depends weakly on the azimuth angle. The standard deviation is 40\,MHz for this case, with 112\,MHz peak--to--peak frequency change. The other two elastic waves have a standard deviation of frequency of about 400\,MHz.  Therefore, the 17.7\,GHz branch has the maximum probability of exciting the stimulated process in the resonator. This prediction supports our experimental observations.
Further analysis shows that polarization of the phonons is nearly collinear with their wave vector in the case of the 17.7\,GHz branch. Hence, the observed SBS process results from the interaction of light and longitudinal sound wave.

In the ideal case of spectrally singular pump and Stokes optical beams, the bulk Brillouin line--center gain $g_b$ can be estimated as (see Eq. 8.3.24 of \cite{boydbook})
\begin{equation}\label{gbulk}
g_b=\frac{\gamma_e^2 \omega^2}{2\pi nV_sc^3\rho \Gamma_{B}},
\end{equation}
where we assume that Stokes and pump wavelengths are equal: $\lambda_{Sj}=\lambda_p = 2\pi c/\omega$. The electrostrictive constant is $\gamma_e = [\rho (\partial \epsilon/\partial \rho)]_S \simeq 0.7$ \cite{mueller35pr}, the fluorite refractive index at $\lambda=1\mu$m is $n=1.429$, the density $\rho=3.18g/cm^3$ and the speed of sound is $V_s = [(C_{11} + C_{12} + 2 C_{44})/(2\rho)]^{0.5}=6.6\times 10^5\ cm/s$. Thus, the value of the bulk Brillouin gain for CaF$_2$  is equal to $g_B=2.8 \times 10^{-9}\ cm/W$  and is two orders of magnitude larger than the bulk Raman gain $g_{r}=2.4 \times 10^{-11}\ cm/W$.

The SBS threshold power for the first Stokes line can be computed similarly to the Raman lasing case:
\begin{equation}
P_{th}=\frac{\pi^2n^2}{g_bQ_pQ_{S1}} \frac{V}{\lambda_p \lambda_{S1}}.
\end{equation}
Selecting realistic values for a 5\,mm cavity, $V=5\times10^{-6}\ cm^3$, $Q_p=Q_{S1}=10^{9}$,
$\lambda_p=\lambda_{S1}=1\ \mu m$, we obtain $P_{th} \simeq 3.6\ \mu W$, which is close to the experimentally observed values.

The first ever observation of an SBS in high-Q whispering gallery mode resonators,to the best of our knowledge, was presented. We provided theoretical analysis verifying our experimental results. The observation of doubly resonant SBS in miniature optical WGM resonators enables ultra--narrow linewidth Brillouin lasers for sensitive optical gyros and stable microwave signal generation. The record high optical Q factor, mechanical stability and robustness of the crystalline WGM resonators make them naturally attractive elements for such application.

Since Brillouin lasing may be detrimental in other applications of crystalline WGM resonators, a method for suppressing SBS would be valuable. Tailoring the resonator spectrum represents one such solution. A single mode fluorite resonator was fabricated with Q factor approaching $10^9$ and diameter of 5\,mm. This cavity had an FSR of around 11\,GHz and no modes were present at the Brillouin offset. No SBS was observed in this cavity.

The research described in this paper was carried out at the Jet Propulsion Laboratory, California Institute of Technology, under a contract with the National Aeronautics and Space Administration, and support from DARPA MTO.

\bibliography{SBS}
\end{document}